\documentclass[aps,prd,reprint,groupedaddress]{revtex4}
\usepackage{amsmath,latexsym,mathrsfs}

\usepackage{amsfonts}
\usepackage{amssymb}
\usepackage[utf8]{inputenc}
\usepackage{graphicx}
\usepackage{url}
\UseRawInputEncoding
\usepackage{breakurl}
\usepackage[breaklinks]{hyperref}
\hypersetup{
    colorlinks=true,
    linkcolor=blue,
    filecolor=magenta,      
    urlcolor=cyan,
}

\begin{document}

\title{Covariant Effective Action for Generalized Proca Theories}%

\author{Sukanta Panda}%
\email[]{sukanta@iiserb.ac.in}

\author{Abbas Altafhussain Tinwala}
\email[]{abbas18@iiserb.ac.in}

\author{Archit Vidyarthi}
\email[]{archit17@iiserb.ac.in}
\affiliation{Department of Physics, Indian Institute of Science Education and Research, Bhopal - 462066, India}

\date{\today}

\begin{abstract}
  We investigate quantum stability of the generalised Proca theories in curved spacetime treating gravity as a dynamical field. To compute the quantum gravitational corrections we evaluate divergent part of the effective action at one loop level. We employ Vilkovisky-DeWitt formalism for this task which gives us a gauge invariant and gauge condition independent effective action. It is shown that the quantum corrections are suppressed by a UV cutoff parametrically higher than the Proca mass if the coupling constants are restricted to lie in a certain range. Furthermore it is shown that the quantum corrections remain suppressed even at scales where classical non-linearities dominate over kinetic terms allowing Vainshtein mechanism to work.
\end{abstract}
\maketitle

\section{Introduction}
    
    General Relativity has established itself as not only one of the most successful but also the most beautiful models for gravitation and cosmology. It has stood many experimental tests across broad scales ranging from Solar System where gravitational field is weak to the binary pulsar system
     where gravitational field is much stronger \cite{tests1,tests2,tests3}. Detection of black holes and gravitational waves emerging from their mergers further adds to the glory of the Einstein's theory. General relativity is one of the two foundations for the Standard Model of big bang cosmology, the other one being the cosmological principle to account for the homogeneity and isotropy of the universe. 
    
    After the discovery of late time cosmic acceleration, the best candidate today (among other modified CDM models) for Standard Model of cosmology is the $\Lambda$CDM Model. In the $\Lambda$CDM model, $\Lambda$ stands for the cosmological constant which acts as dark energy. The theoretical basis that cosmological constant originates from vacuum energy of particle physics has been regarded as one of the worst theoretical predictions as the quantum corrections to the vacuum energy leads to the discrepancy between the predicted and the observed values to be astronomical. Yet it answers reasonably many questions like the existence of CMB,the observed late-time cosmic acceleration and the flat galaxy rotation curves and gravitational lensing of light by galaxy clusters which cannot be accounted for by amount of observed ordinary matter, etc. in a simple manner, albeit introducing two mysterious components: dark matter and dark energy, which have not yet been detected experimentally. Besides these, the model is extended to include a third hypothetical component which is the inflaton field that drives the cosmological inflation.
    
    The problem stated above as well as the fundamental problem of constructing a consistent theory of quantum gravity serves as a motivation to modify gravity both in UV and IR. Modifying gravity almost inevitably leads to addition of extra degrees of freedom. The simplest modification is through a homogeneous time dependent scalar field which respects the cosmological principle as well as give rise to accelerated expansion. Theories that modify gravity at large distances must concur with the irrefutable success of General Relativity in describing gravity at comparatively smaller scale such as Solar System. The problem of adding additional degree of freedom is that it manifests itself into a fifth force which must remain hidden at solar scale to comply with General Relativity. Thus, it seemed initially that such theories must be ruled out. However a screening mechanism was conjectured by Vainshtein \cite{vainshtein} who showed that close to the source general relativistic solutions can be obtained through the effect of higher derivative non-linear interactions of scalar field. There are many theories that modify gravity at large distances. These include massive gravity theories \cite{MG1,MG2,MG3} where scalar field appears as the helicity-0 part of the graviton and the higher dimensional theories like the five dimensional DGP gravity\cite{DGP} where gravity is mediated by massive gravitons from 4D point of view. As briefed above these theories require higher derivative non-linear interactions for the Vainshtein mechanism to work. However this creates the possibility of getting higher than second order equations of motion leading to Ostrogradsky ghost instabilities \cite{og1,og2}. Thus the theories that attempt to modify gravity at large scales must contain only special higher derivative interactions that keeps the pathologies away. It is known that the Horndeski theories\cite{horndeski} are the most general scalar-tensor theory with second order equations of motions (Note: Galileon theory are most general scalar theories with second order equations of motion). These have gained importance because theories like massive gravity and higher dimensional gravity are known to have galileon type interactions in decoupling limit.\cite{IntroVainshtein}.
    
    Modifying GR by adding a scalar degree of freedom is the easiest and most popular since it complies with the observed homogeneity and isotropy of the universe. Thus there's an abounding work on application of Horndeski theories and its generalisations to cosmology\cite{GalCos1,GalCos2,GalCos3,GalCos4,GalCos5,GalCos6,GalCos7}. 
    
    But what about vector fields? Can they bring about viable modifications to GR?  Massless vector field may not be a good idea since it is impossible to construct higher order derivative self interactions to get gauge and Lorentz invariant second order equations of motion of massless vector field. The story is different for massive vector field though. The mass term breaks U(1) gauge symmetry which allows for higher order derivative self interactions with a naturally incorporated Vainshtein mechanism\cite{VainshteinProca}. The ghost free theory of massive vector field with higher order derivative self interactions are known as the generalised Proca theories and have been further extended in curved spacetimes\cite{GP1,GP2,GP3}. These have already found applications to cosmology where the temporal part of proca field, $\phi (t)$, owing to the structure of generalised proca action does not have any dynamics and satisfies algebraic equation making it possible to have a de Sitter solution, serving as a dark energy model. \cite{ProcaCosmo1,ProcaCosmo2,ProcaCosmo3,ProcaCosmo4}.
    
    One issue which is undoubtedly the most unsettling is the possible destabilization of ghost free classical structure under loop corrections. It is not surprising that various quantum corrections will involve higher order derivatives of the vector field. And since it is very much likely this is going to happen then ``what is the scale under which they remain under control?" can be the only next sensible question to ask. Although a thorough analysis to answer the above question had already been undertaken earlier at one loop level(see \cite{QSProca}), it was limited to generalised proca theory in flat spacetime. In this work we attempt to answer this question for generalised proca theory in curved spacetime treating gravity as a dynamical field. 
    
    Computing effective action for this task would be an elegant way to obtain quantum corrections. Indeed, the standard effective action is known to be the generator of 1PI diagrams. However it is not possible to make use of standard effective action if one wants to compute anything but S-matrix elements due to several ambiguities as has been noted in the past \cite{ambiguities1}, where the authors used the standard effective action formalism to explain the smallness of extra dimensions by computing the quantum corrections in the context of Kaluza-Klein theories. There the authors noted the dependence of effective action on the choice of gauge. Another calculation performed \cite{ambiguities2} in the context of Einstein theory coupled to a gauge theory also was found out by \cite{showambiguity} to run into the same problem. Although DeWitt's method of background/mean field gauge improves the formalism in that the effective action is gauge invariant by construction (\cite{Odintsovbook}for details) , the effective action is still dependent on the gauge condition. This problem exists even in non-gauge theories like non-linear $\sigma$ models where there is no preferred coordinate system in which the field can take values. The problem lies in the manifest dependence of standard effective action as well as its improved version (by choosing background field gauge) on parameterization of classical fields, a fact first noted by Vilkovisky \cite{unique}. To overcome these shortcomings Vilkovisky \cite{unique} developed the unique effective action formalism which was further modified by DeWitt\cite{dewitt}. The formalism is now known as the Vilkoviksky-DeWitt effective action (VDEA) formalism which furnishes an effective action, which is gauge invariant as well as gauge condition independent. It is important to note however that although VDEA is perturbatively a sum of one particle irreducible graphs it is not a generator of 1PI n point functions\cite{physinterpvilk}. Recent works such as \cite{features1, features2} highlights important features of VDEA in brief, where as \cite{Tomsbook} gives a detailed description of VDEA and its applications.
    
    Gravity being treated dynamically we are essentially dealing with a gauge theory in this work. Also we wish to keep the background vector field which is not on mass shell in general. Thus we use VDEA formalism to compute an effective action devoid of any gauge ambiguities as briefed above. In light of this, the paper is organised as follows: We start with a brief outline of VDEA and also discuss how to compute it in \hyperref[SecII]{Sec.II}. In \hyperref[SecIII]{Sec.III} we start with a specific form of generalised proca theory in curved spacetime and compute the divergent part of one loop VDEA around a flat spacetime background in \hyperref[SecIV]{Sec.IV}. \hyperref[SecV]{Sec.V} is then dedicated for quantum stability analysis before we sum up our work with concluding remarks in \hyperref[SecVI]{Sec.VI}.

\section{Vilkovisky-DeWitt Effective Action}\label{SecII}
    Effective action is known to be the generator of one-particle irreducible (1PI) diagrams, which reduces to the classical action in the background field limit, and contains all information about the quantum fluctuations in the form of a perturbative expansion where the order of $\hbar$ gives the number of loops involved. We focus our attention in this work only up to one loop level.
    
    Before we proceed to lay an outline of VDEA prescription we familiarise ourselves with the DeWitt notation. The field variables are denoted by $\phi^i$ where any discrete field index as well as spacetime argument of the field has been condensed in the label $i$, i.e $\phi^i = \phi^I(x)$, where $I$ is the conventional field index. For example, if it is a scalar field then $\phi^i = \varphi(x)$, if it is a vector field then $\phi^i = A_\mu(x)$, and if it is a second rank tensor field then $\phi^i = g_{\mu\nu}(x)$. We also adopt the following summation convention in $n$ dimensions:
    \begin{equation}
        \phi^i B_{ij} \phi^j = \int d^nx\int d^nx' \phi^I(x) B_{IJ}(x,x') \phi^J(x')
    \end{equation}
    Also from,
    \begin{align}
        \phi^I(x) &= \int d^nx' |g(x')|^{1/2}\delta(x,x') \phi^I(x')\nonumber\\
                & = \int d^nx' \tilde\delta(x,x') \phi^I(x')
    \end{align}
 we define $\tilde\delta(x,x')$:
    \begin{equation}
        \tilde\delta(x,x') = |g(x')|^{1/2}\delta(x,x')
    \end{equation}
    where $\delta(x,x')$ is the conventional bi-scalar Dirac $\delta$ distribution.
    
    In the light of the above adopted notation the standard effective action $\Gamma[\bar\phi]$ in path integral form is given by
    \begin{equation}\label{sea}
	    	\exp{\frac{i}{\hbar} \Gamma [\bar{\phi}]} = \int [\mathcal{D} \phi] \exp{\frac{i}{\hbar}\left\{S[\phi] -(\phi^i- \bar{\phi}^i)\frac{\delta\Gamma[\bar{\phi}]}{\delta \bar{\phi}^i}\right\}}
    \end{equation}
    In the above expression, $[\mathcal{D} \phi]$ indicates integral to be taken over all field configurations and $\bar\phi^i$ is the background field.
    
    It is straightforward to see that the above expression depends on the choice field parameterization, the fault lying clearly in the use of $\phi^i-\bar\phi^i$, which is not covariant under coordinate changes. This issue was addressed by Vilkovisky who formulated the unique effective action\cite{unique}, later modified by DeWitt\cite{dewitt} resulting in,
     \begin{equation}
         \exp{\frac{i}{\hbar}\Gamma[\bar{\phi};\phi_*]} = \int \prod_id\phi^i |g[\phi]|^\frac{1}{2} \left|\Delta[\phi_*;\phi]\right|
         \exp{\frac{i}{\hbar}\left\{S[\phi] + \frac{\delta\Gamma[\bar{\phi};\phi_*]}{\delta\sigma^i[\phi_*;\bar{\phi}]}\left(\sigma^i[\phi_*;\bar{\phi}] - \sigma^i[\phi_*;\phi]\right)\right\}}
    \end{equation}
    Here, the factor $(\phi^i-\bar{\phi}^i)$ was replaced with using Synge's world function
    \begin{equation*}
        \sigma^i[\phi_*;\phi] = g^{ij}[\phi_*]\frac{\delta}{\delta \phi_*^j}\sigma[\phi_*;\phi]
    \end{equation*}
    $\Delta[\phi_*;\phi]$ is the Van Vleck-Morette determinant, where the choice of $\phi_*$ is arbitrary, and $g[\phi]$ represents the metric of the field configuration space at point $\phi$. DeWitt assigned $\phi_* = \bar{\phi}$, resulting in DeWitt effective action
    \begin{equation}\label{dwea}
	\exp{\frac{i}{\hbar}\Gamma_D[\bar{\phi}]} = \int \prod_id\phi^i |g[\phi]|^\frac{1}{2} \left|\Delta[\bar{\phi};\phi]\right| \exp{\frac{i}{\hbar} \left\{S[\phi] -
	\sigma^i[\bar{\phi};\phi]C^{-1j}_i \frac{\delta\Gamma_D[\bar{\phi}]}{\delta\bar{\phi}^j}\right\}}
	\end{equation}
	where $C^i_j=\left<\sigma^i_{;j}[\bar{\phi};\phi]\right>$, covariant derivatives are taken w.r.t. the field-space connection, and unprimed index in the derivative represents that the derivative is w.r.t. the first argument. Also, $\left<A\right>$ represents the statistical average of some functional $A$ w.r.t. the action. Eq \hyperref[dwea]{6} is really a recursive relation  which can be expanded perturbatively, to arrive at the one-loop effective action(see \cite{Tomsbook} for details),
	\begin{equation}
	    \Gamma^{(1)}[\bar{\phi};\phi_*]=\frac{i\hbar}{2}\ln\det S^{;i}_j = \frac{i\hbar}{2}\ln\det (g^{ij}S_{;ij})
	\end{equation}
    which looks similar to standard effective action, except for a covariant derivative instead of an ordinary derivative w.r.t. fields, and appearance of field space metric.
    
    In case of gauge theories we expect effective action to be gauge invariant as well as independent of gauge condition. For gauge theories it becomes mandatory that off-shell effective action be independent of field parameterization. A simply way to see this is to note the fact that when one employs Fadeev-Popov procedure to convert a gauge action to a non-gauge one by choosing a specific gauge slice, then geometrically this means we are selecting a specific field parameterization. Thus if effective action depends on field parameterization then it will also depend on what gauge choice we make. 
    
    For gauge theories classical action is invariant under gauge transformation:
    \begin{equation}\label{gtran}
        \delta \phi^i = K^i_\alpha \delta \epsilon^\alpha
    \end{equation}
    where $\delta\epsilon^\alpha$ are parameters that characterise the transformations and $\mathbf{K}_\alpha=K^i_\alpha\frac{\delta}{\delta\phi^i}$  are vectors in field-space that act as generators of the transformations. These form a Lie algebra
    \begin{equation}
        \left[\mathbf{K}_\alpha[\phi],\mathbf{K}_\beta[\phi]\right]=-f_{\alpha\beta}^\gamma \mathbf{K}_\gamma[\phi]
    \end{equation}
    where $f_{\alpha\beta}^\gamma$ are the structure constants for the algebra.

    For the covariant derivates one requires the VDW connections which are in general quite complicated. To ease the calculation we choose a special gauge called the Landau-DeWitt gauge\cite{fradkin} which allows us to replace the complicated VDW connections with the standard christoffel connections derived from the metric over field space
    \begin{equation}\label{ldg}
   \chi_\alpha =  K^i_\alpha[\phi_*]g_{ij}[\phi_*](\phi^j-\phi^j_*)=0
    \end{equation}
    
    With this gauge choice one loop effective action for gauge theories turns out to be\cite{Tomsbook,huggins}
    \begin{equation}\label{effac}
        \Gamma[\bar\phi] = -\text{ln det} Q_{\alpha\beta}[\bar\phi] + \dfrac{1}{2}\lim_{\alpha\rightarrow 0}\text{ln det}\Big(S_{;j}^{i} + \dfrac{1}{2\alpha}K^i_\beta[\bar\phi]K^\beta_j[\bar\phi]\Big)
    \end{equation}
    where $Q_{\alpha\beta}$ is the ghost term given by
    \begin{equation}\label{ghost}
        Q^\alpha_\beta =\chi^\alpha_{, \ i}[\phi]K^i_\beta[\phi]
    \end{equation}
    
    \subsection{Computation of Effective Action}\label{SecIIa}
    To compute the above expression one can compute the differential operator corresponding to $S_{;j}^{i} + \dfrac{1}{2\alpha}K^i_\beta[\bar\phi]K^\beta_j[\bar\phi]$. One can then employ heat kernel technique\cite{scholar,Tomsbook} to compute the divergences in effective action. However this technique has limited applications as it can be applied only to some restricted class of operators called minimal operators and can in certain cases be extended to non-minimal ones following the work of Barvinsky \& Vilkovisky\cite{nonminimal}(see \cite{ourpaper,unimodular} for its application). For our work it turns out to be highly complicated to employ this method, so we take a different route. We tweak the expression \hyperref[effac]{(11)} as follows:
    \begin{equation}\label{pert}
        -\dfrac{1}{2}\text{ln det}\Big(S_{;j}^{i} + \dfrac{1}{2\alpha}K^i_\beta[\bar\phi]K^\beta_j[\bar\phi]\Big) = \text{ln}\int [d\psi]\exp\{-S\}
    \end{equation}
    where 
    \begin{align}
    &S = S_q + S_{gb}\nonumber\\
        &S_q = \dfrac{1}{2}\psi^i\psi^j S_{;ij} = \dfrac{1}{2}\eta^i(S_{,ij} - \Gamma^k_{ij}S_{,k})\psi^j\nonumber\\
        &S_{gb} = \dfrac{1}{4\alpha}\psi^i K^\beta_i K_{j\beta}\psi^j    \end{align}
    And $\psi^i$ is the quantum part of the field:
    \begin{equation}
        \phi^i = \bar\phi^i + \psi^i
    \end{equation}
    
    We then expand $S$ in \hyperref[pert]{(13)} in powers of background field. The term with zeroth order in background field $S_0$ is quadratic in quantum field and gives us the propagators. Treating interactions as small we then expand the exponential as:
    \begin{align}\label{pertexp}
        \exp\{-S\} &= \exp\{-(S_0 + S_1 + S_2 + ....)\}\nonumber\\
                &= \exp\{-S_0\}\Big(1 - S_1 + \dfrac{(S_1)^2}{2} - ..\Big)\Big(1 - S_2 + \dfrac{(S_2)^2}{2} - ..\Big)....
    \end{align}
    The expansion can then be worked out up to desired accuracy. In this paper we will obtain the divergences in one loop effective action up to quadratic power of background field. Thus for our work \hyperref[pertexp]{(16)} reduces to
    \begin{equation}\label{pertquad}
    \exp\{-S\} = \exp\{-S_0\}\Big(1 - S_1 - S_2 + \dfrac{(S_1)^2}{2}\Big)
    \end{equation}
     Performing the integral in eq \hyperref[pert]{(11)} we get,
     \begin{align}
         \text{ln}\int [d\psi]\exp\{-S\} &= \text{ln}\int [d\psi]\exp\{-S_0\}\Big(1 - S_1 - S_2 + \dfrac{(S_1)^2}{2}\Big)\nonumber\\
         &=\text{ln}\Big(1 -\langle S_1\rangle - \langle S_2\rangle + \dfrac{\langle (S_1)^2\rangle}{2}\Big)\nonumber\\
         &=-\langle S_1\rangle - \langle S_2\rangle + \dfrac{\langle (S_1)^2\rangle}{2}
     \end{align}
     The angular brackets represents expectation value in path integral formulation which can be computed by employing Wick's theorem. In this light we finally obtain one loop effective action up to quadratic in power of background field:
     \begin{equation}\label{finaleffac}
     \Gamma[\bar\phi] = -\text{ln det} Q_{\alpha\beta} + \lim_{\alpha\rightarrow 0}\Big(\langle S_1\rangle + \langle S_2\rangle - \dfrac{\langle (S_1)^2\rangle}{2}\Big)
     \end{equation}
     
     where $Q_{\alpha\beta}$ is the ghost term defined in \hyperref[ghost]{(12)}
     \\
     
\section{Generalised Proca Action in curved spacetime}\label{SecIII}
    The theory of massless vector field is limited only to the Maxwell kinetic term if we demand Lorentz and gauge invariance of second order equations of motion. However for a massive vector the constraint of U(1) gauge invariance is lifted allowing for higher order derivative interactions. In this regard, the most general theory of massive vector in curved spacetime which gives second order equations of motion for both vector field as well as the metric is given by\cite{GP1,GP2,GP3}
    \begin{align}
        &S = -\int d^4x\sqrt{-g}\big(\mathcal{L}_F + \sum_{i=2}^{6}\mathcal{L}_i\big)\nonumber\\
        &\mathcal{L}_F = -\dfrac{1}{4}F_{\mu\nu}F^{\mu\nu}\nonumber\\
        &\mathcal{L}_2 = G_2(X)\nonumber\\
        &\mathcal{L}_3 = G_3(X)\nabla_\mu A^\mu\nonumber\\
        &\mathcal{L}_4 = G_4(X)R + G_{4,X}((\nabla_\mu A^\mu)^2 + c_2(\nabla_\mu A_\nu \nabla^\mu A^\nu) - (1 + c_2)\nabla_\mu A_\nu \nabla^\nu A^\mu)\nonumber\\
        &\mathcal{L}_5 = G_5(X)G_{\mu\nu}\nabla^\mu A^\nu - \dfrac{1}{6}G_{5,X}((\nabla_\mu A^\mu)^3 -3d_2\nabla_\mu A^\mu \nabla_\rho A_\sigma \nabla^\rho A^\sigma - 3(1-d_2)\nabla_\mu A^\mu \nabla_\rho A_\sigma \nabla^\sigma A^\rho\nonumber\\ & \hspace{8mm}+ (2-3d_2) \nabla_\rho A_\sigma \nabla^\mu A^\rho \nabla^\sigma A_\mu + 3d_2 \nabla_\rho A_\sigma \nabla^\mu A^\rho \nabla_\mu A^\sigma)\nonumber\\
        &\mathcal{L}_6 = G_6(X)\mathcal{L}^{\mu\nu\alpha\beta}\nabla_\mu A_\nu \nabla_\alpha A_\beta + \dfrac{1}{2}G_{6,X}(X)\tilde F^{\alpha\beta}\tilde F^{\mu\nu}\nabla_\alpha A_\mu\nabla_\beta A_\nu\label{GPaction}
        \end{align}
        where 
        \begin{align}
           &X = -\dfrac{1}{2}A^\mu A_\mu \nonumber\\
           &G_{i,X}(X) = \dfrac{\partial G_i(X)}{\partial X}\nonumber\\
           &\mathcal{L}^{\mu\nu\alpha\beta} = \dfrac{1}{4}\epsilon^{\mu\nu\rho\sigma}\epsilon^{\alpha\beta\gamma\delta}R_{\rho\sigma\gamma\delta}\nonumber\\
           &\tilde F^{\mu\nu} = \dfrac{1}{2}\epsilon^{\mu\nu\alpha\beta}F_{\alpha\beta}
        \end{align}
    An overall minus sign in \hyperref[GPaction]{(20)} is only a convention.
    The task of computing effective action is in general highly non trivial for any general form of $G_i(X)$. For our work we choose a specific form of the above action with
    \begin{align}
        &G_2(X) = m^2X\nonumber\\
        &G_3(X) = 2bX = -bA^\mu A_\mu\nonumber\\
        &G_4(X) = c_0 + c_1X = c_0 - \dfrac{1}{2}A^\mu A_\mu
    \end{align}
    We choose $c_0 = 2/\kappa^2$ to get the usual Einstein-Hilbert term in action. We also disregard $\mathcal{L}_5$ and $\mathcal{L}_6$ to make computation less complicated. Thus we are interested in the following subclass of generalised proca theory
    \begin{align}
        S = &\int d^4x\sqrt{-g}\Big(-\dfrac{2}{\kappa^2}R + \dfrac{1}{4}F_{\mu\nu}F^{\mu\nu} + \dfrac{1}{2}m^2A^\mu A_\mu + bA^\mu A_\mu\nabla_\rho A^\rho + \dfrac{1}{2}c_{1R}A^\mu A_\mu R\nonumber\\&\hspace{10mm} - c_1((\nabla_\mu A^\mu)^2 + c_2(\nabla_\mu A_\nu \nabla^\mu A^\nu) - (1 + c_2)\nabla_\mu A_\nu \nabla^\nu A^\mu)\Big)\label{finalGP}
        \end{align}
where $b$, $c_1$ and $c_2$ are just dimensionless constants.

\section{Vilkovisky-DeWitt Effective action for Generalised Proca theory in curved spacetime}\label{SecIV}

In the absence of U(1), the only gauge symmetry of the \hyperref[finalGP]{(23)} is the general coordinate transformation under which
\begin{align}\label{gt}
    &g_{\mu\nu} \rightarrow g_{\mu\nu} - \delta \epsilon^\lambda g_{\mu\nu,\lambda} - g_{\mu\lambda}\delta\epsilon^\lambda_{, \ \nu} - g_{\nu\lambda}\delta\epsilon^\lambda_{, \ \mu}\nonumber\\
   &A_\mu \rightarrow A_\mu - \delta\epsilon^\nu A_{\mu,\nu} - \delta\epsilon^\nu_{, \ \mu}A_\nu
\end{align}
This gives us the form of generators of gauge transformation from \hyperref[gtran]{(8)}
\begin{align}\label{gengtran}
     &K^1_{\mu\nu \ \alpha}(x,x') = -(\partial_\alpha g_{\mu\nu}(x) + g_{\alpha\mu}(x)\partial_\nu + g_{\alpha\nu}(x)\partial_\mu)\tilde{\delta}(x,x')\nonumber\\
     &K^2_{\mu\alpha}(x,x') = -(\partial_\alpha A_\mu(x) + A_\alpha(x)\partial_\mu)\tilde{\delta}(x,x')
\end{align}
where the generators $K^{g_{\mu\nu}}_\alpha$ and $K^{A_{\mu}}_\alpha$ have been simply written as $K^1_{\mu\nu \ \alpha}$ and $K^2_{\mu\alpha}$ respectively by adopting the notation here and throughout the subsequent work: $g_{\mu\nu}\rightarrow 1$, \ \ $A_{\mu}\rightarrow 2$ whenever these appear as field indices. Like for example the field space metric component $G_{g_{\mu\nu}g_{\alpha\beta}}$ is replaced by $(G_{11})^{\mu\nu\alpha\beta}$. .

\subsection{Field Space Metric and Connections}
The central object in the formulation of covariant effective action is the field space metric. To find the field space metric we compare the gauge-fixed action with the line element in field space
\begin{equation}
    ds^2 = \int d^4x d^4x' g_{ij}(x,x')d\phi^i(x)d\phi^j(x')
\end{equation}
However the Landau-DeWitt gauge itself depends upon the field space metric through \hyperref[ldg]{(10)}. From inspection of the action \hyperref[finalGP]{(23)} it is clear from the presence of non-minimal couplings that field space metric isn't going to be diagonal. To proceed further we consider the following ansatz for the field space metric up to quadratic order in vector field
\begin{align}
    &(G_{11})^{\mu\nu\alpha\beta}(x,x') = \big(F_{1}(A)\mathcal{G}^{\mu\nu\alpha\beta} + F_{2}(A)(A^\mu A^\nu g^{\alpha\beta} + A^\alpha A^\beta g^{\mu\nu})\nonumber\\& \hspace{25mm} + F_{3}(A)(A^\mu A^\alpha g^{\nu\beta} + A^\mu A^\beta g^{\alpha\nu} + A^\nu A^\alpha g^{\beta\mu} + A^\nu A^\beta g^{\alpha\mu})\big)_x\sqrt{g(x)}\tilde\delta(x,x')\nonumber\\
    &(G_{12})^{\mu\alpha\beta}(x,x') = \big(H_1(A)(A^\alpha g^{\mu\beta} + A^\beta g^{\alpha\mu}) + H_2(A)A^\mu g^{\alpha\beta}\big)_x\sqrt{g(x)}\tilde\delta(x,x') \nonumber\\
    &(G_{22})^{\mu\nu}(x,x') = \big(J_1(A)g^{\mu\nu} + J_2(A)A^\mu A^\nu\big)_x\sqrt{g(x)}\tilde\delta(x,x')\label{metricansatz}
\end{align}
where a subscript ``$x$" has been attached to indicate the spacetime argument of the functions enclosed in brackets. Here $\mathcal{G}^{\mu\nu\alpha\beta}$ is the Wheeler-DeWitt metric
\begin{equation}\label{WDmetric}
    \mathcal{G}^{\mu\nu\alpha\beta} = g^{\mu(\alpha}g^{\beta)\nu} - \dfrac{1}{2}g^{\mu\nu}g^{\alpha\beta}
\end{equation}
where (..) implies symmetrization of the enclosed indices.

We split the fields as background plus quantum
\begin{align}\label{fieldsplit}
&g_{\mu\nu} = \eta_{\mu\nu} + h_{\mu\nu}\nonumber\\
&A_{\mu} = \bar A_{\mu} + \mathscr{A}_{\mu}
\end{align}
where $\eta_{\mu\nu}$ is the Minkowski metric representing the flat spacetime background and $\bar A_{\mu}$ is the background vector field. $h_{\mu\nu}$ and $\mathscr{A}_{\mu}$ are the quantum fluctuations in gravity and the vector field respectively.

The field space metric gives us the Landau-DeWitt gauge using \hyperref[ldg]{(10)} and \hyperref[gengtran]{(25)}. Here however we just bring in a factor of $1/\sqrt{F_1(A)}$ which does not alter the gauge condition.
\begin{align}
\chi_\lambda(x) &= K^i_\alpha[\phi_*]g_{ij}[\phi_*](\phi^j-\phi^j_*)/\sqrt{F_1(A)}\nonumber\\& = [2\partial^\alpha(h_{\alpha\lambda}F_1(A)) - \partial_\lambda(hF_1(A)) + 2\partial_\alpha( F_2(A) A_\lambda A^\alpha h) + 2\partial_\lambda( F_2(A) A^\alpha A^\beta h_{\alpha\beta})\nonumber\\& + 4\partial_\alpha( F_3(A)A^\alpha A^\beta h_{\lambda\beta}) + 4\partial_\alpha( F_3(A)A_\beta A_\lambda h^{\alpha\beta})+ 2\partial_\alpha( H_1(A) A_\lambda \mathscr{A}^\alpha) \nonumber\\&+ 2\partial_\alpha(H_1(A)A^\alpha\mathscr{A}_\lambda) + 2\partial_\lambda(H_2(A)A^\alpha\mathscr{A}_\alpha) - 2(\partial_\lambda A^\beta) A^\alpha H_1(A)h_{\alpha\beta} - (\partial_\lambda A_\mu) A^\mu h H_1(A) \nonumber\\&+ 2\partial^\beta(H_1(A)A^\alpha A_\lambda h_{\alpha\beta}) + \partial_\alpha(H_1(A) A^\alpha A_\lambda h) - \partial_\lambda A^\alpha J_1(A)\mathscr{A}_\alpha + \partial^\alpha(A_\lambda J_1(A)\mathscr{A}_\alpha)\nonumber\\& - \partial_\lambda A_\alpha J_2(A)A^\alpha A^\beta\mathscr{A}_\beta + \partial_\alpha(A_\lambda J_2(A)A^\beta A^\alpha\mathscr{A}_{\beta})]/\sqrt{F_1(A)}
\end{align}
The gauge breaking action is given by:
\begin{equation}
    S_{gb} = \dfrac{1}{4}\int d^4x \sqrt{-g(x)}\chi_\alpha(x)\chi^\alpha(x)
\end{equation}
To obtain the field space metric we expand the action \hyperref[finalGP]{(23)} in quadratic power of the quantum fields. Only specific terms will contribute to field space metric. These go as:
\begin{align}\label{forms}
    &\partial_\mu h \partial^\mu h\nonumber\\
    &\partial_\mu h \partial^\mu \mathscr{A}\nonumber\\
    &\partial_\mu \mathscr{A} \partial^\mu \mathscr{A}
\end{align}
The gauge fixed lagrangian $\mathcal{L}_{gf}$ gives
\begin{align}
   \mathcal{L}_{gf}& = -\Big[\Big(\dfrac{1}{2}g^{\alpha\mu}g^{\beta\nu}+\dfrac{1}{2}g^{\alpha\nu}g^{\beta\mu}\Big)\Big(\dfrac{1}{4}A^2c_1- 1\Big) + g^{\alpha\beta}g^{\mu\nu}\Big(1-\dfrac{1}{4}A^2c_1-\dfrac{F_1(A)}{2}\Big) \nonumber\\&\hspace{18mm}+  A^\mu A^\nu g^{\alpha\beta}(\dfrac{1}{2}c_1+F_2(A)) +  A^\alpha A^\beta g^{\mu\nu}(\dfrac{1}{2}c_1+F_2(A))\nonumber\\&\hspace{18mm}-\dfrac{1}{4}c_1(A^\alpha A^\mu g^{\beta\nu} + A^\alpha A^\nu g^{\beta\mu} + A^\beta A^\mu g^{\alpha\nu} + A^\beta A^\nu g^{\alpha\mu})\Big]\partial^{\lambda}h_{\alpha\beta}\partial_{\lambda}h_{\mu\nu}\nonumber\\
     &\hspace{8mm}-2 \Big[\dfrac{1}{2}c_1 A^\alpha g^{\mu\beta} + \dfrac{1}{2}c_1 A^\beta g^{\alpha\mu} - c_1 A^\mu g^{\alpha\beta}+H_2(A)A^\mu g^{\alpha\beta}\Big]\partial^{\lambda}h_{\alpha\beta}\partial_{\lambda}\mathscr{A}_{\mu}\nonumber\\
     &\hspace{8mm}+ \Big[(1 - 2c_1c_2)g^{\mu\nu}+2\dfrac{(H_2(A))^2}{F_1(A)}\mathscr{A}^\mu\mathscr{A}^\nu\Big]\partial^{\lambda}\mathscr{A}_{\mu}\partial_{\lambda}\mathscr{A}_{\nu} + (...)
\end{align}
Here (...) represents all other terms not of the form \hyperref[forms]{(32)}. We compare the above expression with line element
\begin{equation}
    ds^2 = \int G_{ij}(\phi)d\phi^id\phi^j = \int G_{11}^{\alpha\beta\mu\nu}dh_{\alpha\beta}dh_{\mu\nu} + 2G_{12}^{\alpha\mu}dh_{\alpha\beta}d\mathscr{A}_{\mu} + G_{22}^{\mu\nu}d\mathscr{A}_\mu d\mathscr{A}_\nu
\end{equation}
Through this we obtain
\begin{align}
    &F_1(A) = 1-\dfrac{1}{4}A^2c_1\nonumber\\
    &F_2(A) = -\dfrac{c_1}{4}\nonumber\\
    &F_3(A) = \dfrac{c_1}{4}\nonumber\\
    &H_1(A) = -\dfrac{c_1}{2}\nonumber\\
    &H_2(A) = \dfrac{c_1}{2}\nonumber\\
    &J_1(A) = 1-2c_1c_2\nonumber\\
    &J_2(A) = \dfrac{2(H_2(A))^2}{F_1(A)} = \dfrac{c_1^2}{2-\dfrac{1}{2}A^2c_1}
\end{align}
The connections formed from the field space metric are given by
\begin{equation}
    \Gamma^i_{jk} = \dfrac{1}{2}g^{il}(g_{lj,k} + g_{lk,j} - g_{jk,l})
\end{equation}
These are quite complicated and have been presented in the Appendix \hyperref[AppA]{A} along with the inverse field space metric.
\subsection{Effective action}
For typographical convenience we will omit the bar over background fields from now on. Most of the intermediate expressions in the computation of effective action involve an immense number of terms to be presented here, we therefore present only the most relevant ones.

The term with zeroth power of background field reads
\begin{align}
    S_0 = \int d^4x\Big(&\dfrac{1}{2}m^2\mathscr{A}_\mu \mathscr{A}^\mu - c_1\partial_\mu\mathscr{A}^\mu\partial_\nu \mathscr{A}^\nu +\Big(c_1c_2 + c_2 -\dfrac{1}{2}\Big)\partial_\mu \mathscr{A}_\nu\partial^\nu\mathscr{A}^\mu+ \Big(\dfrac{1}{2} - c_1\Big)\partial_\mu\mathscr{A}_\nu \partial^\mu\mathscr{A}^\nu \nonumber\\&+ \Big(\dfrac{1}{\alpha}-1\Big)\Big(\partial^\mu h_{\mu\nu} - \dfrac{1}{2}\partial_\nu h\Big)^2 - \dfrac{1}{2}h_{\mu\nu}\Box h^{\mu\nu} + \dfrac{1}{4}h\Box h\Big)
\end{align}
This gives us the propagators,
\begin{align}\label{psp}
    &\langle\mathscr{A}_\mu(x)\mathscr{A}_\nu(x')\rangle = D_{1\mu\nu}(x,x') = \int d^4k e^{ik(x-x')}D_{1\mu\nu}(k) \nonumber\\
    &\langle h_{\mu\nu}(x)h_{\alpha\beta}(x')\rangle = D_{2\mu\nu\alpha\beta}(x,x') = \int d^4k e^{ik(x-x')}D_{2\mu\nu\alpha\beta}(k) 
\end{align}
which in momentum space reads,
\begin{align}\label{msp}
   &D_{1\mu\nu}(k) = m^2\eta_{\mu\nu} + (1-2c_2c_2)k_\mu k_\nu\nonumber\\ &D_{2\mu\nu\alpha\beta}(k) = \dfrac{2\eta_{\mu(\alpha}\eta_{\beta)\nu} - \eta_{\mu\nu}\eta_{\alpha\beta}}{2k^2} + \dfrac{(\alpha -1)(\eta_{\mu\alpha}k_\nu k_\beta + \eta_{\mu\beta}k_\nu k_\alpha+ \eta_{\nu\alpha}k_\mu k_\beta + \eta_{\nu\beta}k_\mu k_\alpha)}{2k^4}
\end{align}
    \begin{enumerate}
        \item  \ \ $\langle S_i \rangle$ \\\\
        Using \hyperref[psp]{(38)} one can write down expressions for any general form
\begin{equation}\label{twopointd}
   \int d^4x \int d^4x'\langle(\partial)^m \phi_i(x) (\partial)^n \phi_i(x')\rangle = \int d^4x \int d^4x'\int d^4k(ik)^m(ik)^nD_i(k)e^{ik(x-x')}
\end{equation}
    We encounter terms of the above form in $S_1$ and $S_2$ with coincident spacetime arguments and we use dirac delta function to write them as
    \begin{equation}\label{twopoints}
        \int d^4x f(x)\langle(\partial)^m \phi_i(x) (\partial)^n \phi_i(x)\rangle = \int d^4x f(x)\int d^4x' \delta(x,x') \langle(\partial)^m \phi_i(x) (\partial)^n \phi_i(x')\rangle
    \end{equation}
    where $f(x)$ is any function dependent on parameters and background fields. One then proceeds with the evaluation of the loop integrals using \hyperref[twopointd]{(40)}. Here we will be concerned with the divergent part only and we will employ dimensional regularisation. We have listed the divergent part of relevant loop integrals in the Appendix \hyperref[AppB]{B}. Note that any term which goes like $\langle(\partial)^m h (\partial)^n \mathscr{A}\rangle$ vanishes.
        Due to cubic interaction present in the action $\langle S_1 \rangle$ will not vanish. In momentum space we encounter the following types of loop integrals in $\langle S_1 \rangle$ and $\langle S_2 \rangle$
        \begin{equation}
            \int d^4k \dfrac{k_{\mu_1} .... k_{\mu_n}}{m^2 + (1-2c_1c_2)k^2}
        \end{equation}
     Out of this, integrals involving odd number of $k$'s in the numerator of the integrand will not contribute to the divergent part.
     \item   \ \ $\langle S_1^2 \rangle$
     
     The number of terms in $S_1^2$ is immense. However all the terms appearing here are of the general form as given below.
     \begin{align}\label{s12terms}
       &\int d^4x\int d^4x' f(x)g(x')\langle (\partial)^i\mathscr{A}_\mu(x)(\partial)^j\mathscr{A}_\alpha(x')(\partial)^m\mathscr{A}_\nu(x)(\partial)^n\mathscr{A}_\beta(x')\rangle \nonumber\\
       &\int d^4x\int d^4x' f(x)g(x')\langle(\partial)^i\mathscr{A}_\mu(x)(\partial)^mh_{\rho\sigma}(x)(\partial)^j\mathscr{A}_\nu(x')(\partial)^nh_{\lambda\tau}(x')\rangle\nonumber\\
       &\int d^4x\int d^4x' f(x)g(x')\langle(\partial)^i\mathscr{A}_\mu(x)(\partial)^j\mathscr{A}_\nu(x)(\partial)^mh_{\rho\sigma}(x')(\partial)^nh_{\lambda\tau}(x')\rangle\nonumber\\
       &\int d^4x\int d^4x' f(x)g(x')\langle(\partial)^i\mathscr{A}_\mu(x)(\partial)^j\mathscr{A}_\nu(x)(\partial)^m\mathscr{A}_{\rho}(x')(\partial)^nh_{\lambda\tau}(x')\rangle
    \end{align}
    The first two terms above can be written as
     \begin{align}\label{4a}
         \langle (\partial)^i\mathscr{A}_\mu(x)(\partial)^m\mathscr{A}_\nu(x)(\partial)^j\mathscr{A}_\alpha(x')(\partial)^n\mathscr{A}_\beta(x')\rangle &= \langle (\partial)^i\mathscr{A}_\mu(x)(\partial)^m\mathscr{A}_\nu(x)\rangle\langle(\partial)^j\mathscr{A}_\alpha(x')(\partial)^n\mathscr{A}_\beta(x')\rangle \nonumber\\&+ \langle (\partial)^i\mathscr{A}_\mu(x)(\partial)^j\mathscr{A}_\alpha(x')\rangle\langle(\partial)^m\mathscr{A}_\nu(x)(\partial)^n\mathscr{A}_\beta(x')\rangle \nonumber\\&+ \langle (\partial)^i\mathscr{A}_\mu(x)(\partial)^n\mathscr{A}_\beta(x')\rangle\langle(\partial)^m\mathscr{A}_\nu(x)(\partial)^j\mathscr{A}_\alpha(x')\rangle
     \end{align}
     \begin{align}\label{2a2h}
         \langle(\partial)^i\mathscr{A}_\mu(x)(\partial)^mh_{\rho\sigma}(x)(\partial)^j\mathscr{A}_\nu(x')(\partial)^nh_{\lambda\tau}(x')\rangle &= \langle(\partial)^i\mathscr{A}_\mu(x)(\partial)^mh_{\rho\sigma}(x)\rangle\langle(\partial)^j\mathscr{A}_\nu(x')(\partial)^nh_{\lambda\tau}(x')\rangle \nonumber\\& + \langle(\partial)^i\mathscr{A}_\mu(x)(\partial)^nh_{\lambda\tau}(x')\rangle\langle (\partial)^mh_{\rho\sigma}(x)(\partial)^j\mathscr{A}_\nu(x')\rangle \nonumber\\& + \langle(\partial)^i\mathscr{A}_\mu(x)(\partial)^j\mathscr{A}_\nu(x')\rangle\langle(\partial)^mh_{\rho\sigma}(x)(\partial)^nh_{\lambda\tau}(x')\rangle
     \end{align}
    where we have used Wick's theorem to write four point correlators in terms of sum of products of two point correlators. The first term in RHS of \hyperref[4a]{(44)} does not contribute since it corresponds to disconnected tadpoles. Also first and second term of \hyperref[2a2h]{(45)} vanishes since $\langle(\partial)^m h (\partial)^n \mathscr{A}\rangle$ vanishes. Using these arguments other four point correlators in \hyperref[s12terms]{(43)} vanishes.
    
    The terms that contribute can be Fourier transformed as
    \begin{align}
      \int d^4x &\int d^4x' f(x)g(x')\langle (\partial)^i\mathscr{A}_\mu(x)(\partial)^j\mathscr{A}_\alpha(x')\rangle\langle(\partial)^m\mathscr{A}_\nu(x)(\partial)^n\mathscr{A}_\beta(x')\rangle\nonumber\\ &=  \int d^4x \int d^4x' \int \dfrac{d^4k}{2\pi^4}\int \dfrac{d^4k'}{2\pi^4}\int \dfrac{d^4k''}{2\pi^4}\int \dfrac{d^4k'''}{2\pi^4} e^{ikx}e^{ik'x'}e^{ik''(x-x')}e^{ik'''(x-x')} \  \ \times \nonumber\\&\hspace{16mm}\tilde f(k)\tilde g(k') (ik'')^i(-ik'')^j(ik''')^m(-ik''')^n D_{1\mu\alpha}(k'')D_{1\nu\beta}(k''')\nonumber\\
      &=\int d^4x f(x)\int \dfrac{d^4k}{2\pi^4}\tilde g(k)e^{ikx}\int \dfrac{d^4p}{2\pi^4}(ip)^i(-ip)^j(i(k-p))^m(-i(k-p))^nD_{1\mu\alpha}(p)D_{1\nu\beta}((k-p))
    \end{align}
    A similar expression can be derived from $\int d^4x\int d^4x'\langle(\partial)^i\mathscr{A}_\mu(x)(\partial)^j\mathscr{A}_\nu(x')\rangle\langle(\partial)^mh_{\rho\sigma}(x)(\partial)^nh_{\lambda\tau}(x')\rangle$
    
    These give us the following type of loop integrals which have been given in Appendix \hyperref[AppB]{B}
    \begin{align}
    &\int d^4p \dfrac{p_{\mu_1} \  ... \  p_{\mu_n}}{(m^2+(1-2c_1c_2)p^2)(m^2 + (1-2c_1c_2)(k-p)^2)}\nonumber\\
    &\int d^4p \dfrac{p_{\mu_1} \  ... \  p_{\mu_n}}{p^2(m^2 + (1-2c_1c_2)(k-p)^2)}\nonumber\\
    &\int d^4p \dfrac{p_{\mu_1} \  ... \  p_{\mu_n}}{p^4(m^2 + (1-2c_1c_2)(k-p)^2)}
    \end{align}
    \end{enumerate}

Finally we have to compute the ghost term in \hyperref[finaleffac]{(19)} using \hyperref[ghost]{(12)}. However it turns out that it does not contribute up to quadratic power of background vector field. In light of this, using the procedure outlined above we end up with an immense number of terms which eventually combine to give us the following structure of divergences in one loop effective action \hyperref[effac]{(11)},
\begin{equation}\label{diveffac1}
    \Gamma^{(1)}_{div} = f_1A^\mu A_\mu + f_2\partial_\mu A^\mu + f_3A^\mu\Box A_\mu + f_4 A^\mu\partial_\nu\partial_\mu A^\nu + f_5 A^\mu\Box^2A_\mu + f_6A^\mu\partial_\nu\partial_\mu\Box A^\nu + f_7A^\mu\Box^3A_\mu + f_8A^\mu\partial_\mu\partial_\nu\Box^2A^\nu 
\end{equation}
The form of functions $f_i$ are such that it allows the above expression to be written as follows 
\begin{align}\label{diveffac2}
    \Gamma^{(1)}_{div} = & \tilde g_1A^\mu A_\mu + \tilde g_2\partial_\mu A^\mu + \tilde g_3A^\mu\Box A_\mu + \tilde g_4 A^\mu\partial_\nu\partial_\mu A^\nu + \tilde g_5 A^\mu\Box^2A_\mu + \tilde g_6A^\mu\partial_\nu\partial_\mu\Box A^\nu + \tilde g_7A^\mu\Box^3A_\mu + \tilde g_8A^\mu\partial_\mu\partial_\nu\Box^2A^\nu \nonumber\\&+ g_1F_{\mu\nu}F^{\mu\nu} + g_2F_{\mu\nu}\Box F^{\mu\nu} + g_3F_{\mu\nu}\Box^2F^{\mu\nu}
\end{align}
As one expects there are no terms in the above expression which blow up in the limit $\alpha \rightarrow 0$. The functions $f_i$, $\tilde g_i$ and $g_i$ depend only upon the constant parameters of the theory and are given in Appendix \hyperref[AppC]{C}. We note the following features which we find to be somewhat in agreement with \cite{QSProca} even under inclusion of gravitational corrections.
\begin{enumerate}
    \item Terms involving inverse powers of proca mass are troublesome since they can potentially renormalise classical structure in the limit of vanishing proca mass.
    \item $g_i$ are functions involving either $c_1$, $C (= 1-2c_1c_2)$ or $m$. Thus gauge invariant structures do not arise from $\mathcal{L}_3$.
    \item The term $\partial_\mu A^\mu$ can of course be dropped since it is a total derivative.
\end{enumerate}
\section{Quantum Stability}\label{SecV}
\subsection{Decoupling limit analysis}\label{SecVa}
    Having computed the divergences in one loop effective action we are ready to answer the question whether these corrections introduce any pathology in the classical theory or not. A natural way to answer this question is to see what happens in the limit of vanishing Proca mass. We employ the Stueckelberg formulation by introducing an auxiliary field $\phi$,
    \begin{equation}\label{STform}
        A_\mu\rightarrow A_\mu + \dfrac{\partial_\mu\phi}{m}
    \end{equation}
    This does not affect the number of degrees of freedom propagated by the theory since we have just artificially introduced a redundancy here, i.e. the action is now invariant under
    \begin{equation}\label{gsym}
        A_\mu\rightarrow A_\mu - \partial_\mu\psi \ \ \ \ \ \ \phi\rightarrow \phi + m\psi
    \end{equation}\label{stbergb}
    
    It is clear that in the unitary gauge $\psi = -\phi/m$ we get back the original action. In the limit of vanishing proca mass however \hyperref[gsym]{(51)} is no longer a symmetry as we will see later when we apply the decoupling limit.
    
    To see how metric couples to $A_\mu$ and $\phi$ we expand the action around flat spacetime 
\begin{equation}
    g_{\mu\nu} = \eta_{\mu\nu} + \kappa h_{\mu\nu}
\end{equation}   
where the factor $\kappa = 2/M_p$ canonincally normalizes $h_{\mu\nu}$ with mass dimension one.

The Einstein-Hilbert term coming from $\mathcal{L}_4$ in the action gives free quadratic action for $h_{\mu\nu}$
\begin{equation}
    \mathcal{L}^{(2)}_{EH} = -\dfrac{1}{2}h_{\mu\nu}\mathcal{K}^{\mu\nu\alpha\beta}h_{\alpha\beta}
\end{equation}
    where $\mathcal{K}^{\mu\nu\alpha\beta}$ is dimension-2 operator given by
    \begin{equation}
        \mathcal{K}^{\mu\nu\alpha\beta} = (\eta^{\mu(\alpha}\eta^{\beta)\nu} - \eta^{\mu\nu}\eta^{\alpha\beta})\partial^2 + \eta^{\mu\nu}\partial^\alpha\partial^\beta + \eta^{\alpha\beta}\partial^\mu\partial^\nu - \eta^{\mu(\alpha}\partial^{\beta)}\partial^\nu - \eta^{\nu(\alpha}\partial^{\beta)}\partial^\mu
    \end{equation}

The kinetic term $\dfrac{1}{4}F_{\mu\nu}F^{\mu\nu}$ gives up to first order in $\kappa$
\begin{align}\label{l21}
&\mathcal{L}^{(1)}_2 = \mathcal{L}_{2\phi} + \mathcal{L}_{2A} + \mathcal{L}_{2A\phi}\nonumber\\
&\mathcal{L}_{2\phi} = \dfrac{1}{2}\partial_\mu\phi\partial^\mu\phi + \dfrac{\kappa h_{\mu\nu}}{2}\Big(\dfrac{g^{\mu\nu}\partial_\alpha\phi\partial^\alpha\phi}{2} - \partial^\mu\phi\partial^\nu\phi\Big)\nonumber\\
&\mathcal{L}_{2A} = \dfrac{1}{4}F_{\mu\nu}F^{\mu\nu} + \dfrac{1}{2}m^2A^2 + \kappa h_{\mu\nu}\Big(-\dfrac{1}{2}m^2A^\mu A^\nu
 + \dfrac{1}{4}g^{\mu\nu}\Big(m^2A^2 +\dfrac{1}{2}F_{\mu\nu}F^{\mu\nu}\Big)\nonumber\\ &\hspace{58mm} -\dfrac{1}{2}\partial_\alpha A^\mu\partial^\alpha A^\nu - \dfrac{1}{2}\partial^\mu A_\alpha\partial^\nu A^\alpha + \partial_\alpha A^\mu\partial^\nu A^\alpha\Big)\nonumber\\
&\mathcal{L}_{2A\phi} = mA^\mu\partial_\mu\phi + \kappa h_{\mu\nu}\Big(-mA^\mu\partial^\nu\phi + \dfrac{1}{2}mg^{\mu\nu}A^\alpha\partial_\alpha\phi\Big)
 \end{align}

Lagrangian $\mathcal{L}_3$ up to first order in $\kappa$ reads
 \begin{align}\label{l31}
     \mathcal{L}^{(1)}_3 = &b\partial_\mu A^\mu + \dfrac{b}{m}(2\partial_\nu A^\nu A^\mu\partial^\mu\phi + A^2\partial_\mu\phi\partial^\mu\phi) + \dfrac{b}{m^2}(2A^\mu
    \partial_\mu\phi\partial^2\phi + \partial_\mu A^\mu(\partial\phi)^2)+\dfrac{b}{m^3}\partial_\mu\phi\partial^\mu\phi\partial^2\phi\nonumber\\
    & + \kappa b h_{\mu\nu}X_{31}^{\mu\nu} +\dfrac{\kappa b}{m}h_{\mu\nu}X_{32}^{\mu\nu}+\dfrac{\kappa b}{m^2}h_{\mu\nu}X_{33}^{\mu\nu} + \dfrac{\kappa b}{m^3}X_{34}^{\mu\nu}
\end{align}
We really do not need to know the precise form of the expressions for $X_{31}^{\mu\nu}$, $X_{32}^{\mu\nu}$ and $X_{33}^{\mu\nu}$ since they depend on both $\phi$ and $A$ and we will see it briefly that the terms involving them will vanish in decoupling limit. $X_{34}^{\mu\nu}$ however is purely dependent upon $\phi$ and the coupling between $h$ and $\phi$ is such that it could not be diagonalized by any local redefinition of $h_{\mu\nu}$.

Lagrangian $\mathcal{L}_4$ without the Einstein Hilbert term reads up to first order in $\kappa$ 
\begin{align}\label{l41}
    \mathcal{L}^{(1)}_4 = -\dfrac{c_1c_2}{2}F_{\mu\nu}F^{\mu\nu} + \kappa c_1 h_{\mu\nu}X_{41}^{\mu\nu} + \dfrac{\kappa}{m} c_1h_{\mu\nu}X_{42}^{\mu\nu} + \dfrac{\kappa}{2m^2}c_1h_{\mu\nu}\mathcal{K}^{\mu\nu\alpha\beta}\partial_\alpha\phi\partial_\beta\phi
\end{align}
In the above expression $X_{41}^{\mu\nu}$ and $X_{42}^{\mu\nu}$ depend upon both $A$ and $\phi$. Note that the last term in RHS of \hyperref[l41]{(57)} can be diagonalized by local redefiniton of $h_{\mu\nu}$.

We assume that coupling constants $b$ and $c_1$ are very small which is reasonable for an EFT and of the same order. From \hyperref[l21]{(55)}, \hyperref[l31]{(56)} and \hyperref[l41]{(57)} it is clear that the most dangerous interaction has coefficient $b/m^3$ such that the theory completely loses its predictive power in the limit of vanishing proca mass if $b \sim \mathcal{O}(1)$. Also a free theory admits an effective description at a scale lower than Planck mass. This tells us that $b$ must lie within\cite{dcpl1}
\begin{equation}\label{br}
    \mathcal{O}(m^3\kappa^3) < b < \mathcal{O}(1)
\end{equation}
Note that the term $\dfrac{b}{m^3}\partial_\mu\phi\partial^\mu\phi\partial^2\phi$ having $m^3$ in the denominator has a tendency to be badly divergent in the limit of vanishing mass. Thus there is a tremendous pressure on $b$ to take value as small as possible to maximise the predictive power of EFT. To get the minimum value we look at divergences in one loop effective action. From \hyperref[diveffac2]{(49)} the most dangerous term which could potentially renormalize the classical action is:
\begin{equation}\label{danger}
   \sim \dfrac{\kappa^2}{m^2}(A^\mu\partial_\mu\partial_\nu\partial^2A^\nu - A^\mu\partial^3 A_\mu) \sim \dfrac{\kappa^2}{m^2}F_{\mu\nu}\partial^2F^{\mu\nu} \xrightarrow{DL} ~\dfrac{\kappa^2}{m^2}F_{\mu\nu}\partial^2F^{\mu\nu} 
\end{equation}
Note that we have not yet defined the decoupling limit though but it is expected that a gauge invariant combination of vector field would remain unaffected.

If we take the dimensionless coupling constants to be
\begin{equation}
    c_1 \sim b \sim \kappa^\alpha m^\alpha
\end{equation}
which gives us the energy scale at which the cubic interaction in $\mathcal{L}^{(1)}_3$ becomes important
\begin{equation}\label{es1}
    E^3 = \dfrac{m^3}{\kappa^\alpha m^\alpha} = \dfrac{m^{3-\alpha}}{\kappa^\alpha}
\end{equation}
then the factor $\kappa^2/m^2$ appearing in \hyperref[danger]{(59)} can be written as
\begin{equation}\label{dangerousterm}
    \dfrac{\kappa^2}{m^2} = m^{\tfrac{(3-\alpha)(2-\alpha)}{\alpha} - \alpha + 1}\dfrac{1}{E^{(6/\alpha)}}
\end{equation}
$\alpha$ cannot of course be equal to or less than zero, otherwise $b \geq \mathcal{O}(1)$. To keep the dangerous term \hyperref[dangerousterm]{(59)} under control it must be suppressed by energy scale $E$. So we must have in the limit of vanishing mass
\begin{equation}
    \dfrac{(3-\alpha)(2-\alpha)}{\alpha} -\alpha +1 \geq 0
\end{equation}
Above condition is satisfied for 
\begin{equation}\label{alphacond}
0\leq \alpha \leq 3/2
\end{equation}
The best choice we can make is taking the largest value of $\alpha$ so that $b$ is smallest. Thus, for $\alpha = 3/2$ we get the energy scale from \hyperref[es1]{(61)}
\begin{equation}\label{es}
    E = \sqrt{\dfrac{m}{\kappa}} = \sqrt{mM_P}
\end{equation}
and the decoupling limit
\begin{equation}\label{dl1}
    \kappa \rightarrow 0 \ \ \ \ \ \ m\rightarrow 0 \ \ \ \ \ \ E= \sqrt{\dfrac{m}{\kappa}} \rightarrow \text{constant} 
\end{equation}
Note that the term $\tilde g_8 A^\mu\partial_\mu\partial_\nu\Box^2A^\nu$ which looks intimidating is actually suppressed by scale $E$
\begin{equation}
    \tilde g_8A^\mu\partial_\mu\partial_\nu\Box^2A^\nu \sim \dfrac{b^2}{m^4}A^\mu\partial_\mu\partial_\nu\Box^2A^\nu \xrightarrow{DL} \dfrac{b^2}{m^4}\dfrac{1}{m^2}\partial^\mu\phi\partial_\mu\partial_\nu\Box^2\partial^\nu\phi = \dfrac{1}{E^6}\partial^\mu\phi\partial_\mu\partial_\nu\Box^2\partial^\nu\phi
\end{equation}
In the decoupling limit \hyperref[dl1]{(66)} all but the kinetic terms and cubic pure scalar coupling in $\mathcal{L}_3$ vanishes and the lagrangian reduces to
\begin{equation}\label{declag}
    \mathcal{L}_{DL} = \dfrac{1}{4}F_{\mu\nu}F^{\mu\nu} + \dfrac{1}{2}\partial_\mu \phi\partial^\mu\phi - \dfrac{1}{2}h_{\mu\nu}\mathcal{K}^{\mu\nu\alpha\beta}h_{\alpha\beta} + \dfrac{1}{E^3}\partial_\mu\phi\partial^\mu\phi\partial^2\phi
\end{equation}
It is clear that the above action no longer possesses symmetry \hyperref[gsym]{(51)}. Apart from diffeomorphisms the new symmetry here is the U(1) gauge invariance of $A_\mu$ and the global shift symmetry of $\phi$ indicating that the transverse modes, the longitudinal mode and $h_{\mu\nu}$ have all completely decoupled from each other,
\begin{equation}
    A_\mu \rightarrow A_\mu + \partial_\mu\psi \ \ \ \ \ \ \phi \rightarrow \phi + C
\end{equation}
It is straightforward that all the divergences appearing in the one loop effective action \hyperref[diveffac1]{(49)} are all either suppressed by scale $E$ or vanish when decoupling limit is taken. The ghost free classical structure is thus protected from destabilization arising from quantum corrections up to quadratic in order of the background vector field if the dimensionless coupling constants $b$ and $c_1$ are restricted to lie in the range
\begin{equation}\label{bbound}
\mathcal{O}(\sqrt{m^3\kappa^3}) \leq b,c_1 \leq \mathcal{O}(1)
\end{equation}
\subsection{Hierarchy among classical and quantum operators}
Theories that modify gravity on large scales requires Vainshtein screening that relies on the relative dominance of non-linear terms in classical action compared to linear ones. What concerns us are the quantum induced operators. Is the EFT protected from them at scales where classical non-linear terms become dominant?. To answer this question we simply sum up results from decoupling limit analysis. Let us write down the action including one loop corrections in the decoupling limit.
\begin{equation}\label{qcaction}
\mathcal{L}_{QC} \sim F_{\mu\nu}F^{\mu\nu} + \partial_\mu \phi\partial^\mu\phi - h_{\mu\nu}\mathcal{K}^{\mu\nu\alpha\beta}h_{\alpha\beta} + \dfrac{1}{E^3}\partial_\mu\phi\partial^\mu\phi\partial^2\phi + \dfrac{1}{E^4}F_{\mu\nu}\Box^2F^{\mu\nu} + \dfrac{1}{E^6}\partial_\mu\phi\Box^3\partial^\mu\phi
 \end{equation}
 The first three terms above are kinetic terms. The fourth term is the only surviving classical non-linear term in decoupling limit \hyperref[dl1]{(66)} with the last two terms arising from quantum loops.
 To get a concise picture of how classical non-linearities and quantum induced terms compare with the kinetic terms it is helpful to establish an hierarchy by identifying the classical and quantum expansion parameters[\cite{olDGP,multifieldgal,QSProca}]. The quantum corrected action \hyperref[qcaction]{(71)} can be written succinctly (by suppressing index notation) as
 \begin{equation}
   \mathcal{L}_{QC} \sim  F^2 + (\partial\phi)^2 - H^2 + \alpha_{cl}(\partial\phi)^2 + \alpha_q^2F^2 + \alpha_q^3(\partial\phi)^2
 \end{equation}
 where 
 \begin{equation}\label{param}
 \alpha_{cl} = \dfrac{\partial^2\phi}{E^3} \ \ \ \  \ \ \ \ \ \ \ \ \ \ \ \ \ \alpha_q = \dfrac{\partial^2}{E^2}
 \end{equation}
 are the classical and quantum expansion parameters respectively.
 To discuss Vainshtein mechanism we only need the lagrangian of linearised gravity and $\phi$ since there are no non linear terms in vector field that survive in the decoupling limit.
 
 It is known that Vainshtein screening mechanism comes into play below a distance $R_V$ called vainshtein radius from the source. This happens when the classical non-linear terms dominate over kinetic terms, in other words when $\alpha_{cl} >> 1$, i.e
 \begin{equation}\label{aclcond}
 \dfrac{\partial^2\phi}{E^3} >> 1
 \end{equation}
 A simple way to derive expression for $R_V$ is by demanding the above condition \hyperref[aclcond]{(74)} to hold and arguing that GR receives modification $\phi \sim h$ at distances larger then $R_V$. If we look at the spherically symmetric solutions then linearised equations for $h$ gives us (see \cite{IntroVainshtein} for details)
 \begin{equation}\label{beyvain}
     \phi \sim h = -\dfrac{M_pR_S}{r}
 \end{equation}
 where $R_S$ is the Schwarzschild radius. Using this we find from \hyperref[aclcond]{(74)} that Vainshtein mechanism switches on when,
 \begin{equation}\label{vrad}
     r << \dfrac{({M_PR_S})^{1/3}}{E} = \dfrac{1}{E}\Big(\dfrac{M}{M_P}\Big)^{1/3} = R_V
 \end{equation}
However are we sure if the quantum induced operators does not spoil this property. This issue has already been resolved in the context of galileon theories in \cite{multifieldgal} where the authors also end up with the same cubic interaction term owing to the fact that Proca theories end up in galileon terms in DL. In this light, we only brief the work of \cite{multifieldgal}(see secII) here,
\begin{align}
    &\alpha_{cl} << 1, \ \ \ \ \ \ \alpha_q <<1 \ \ \ \ \ \ \ \ \text{when} \ \ \ \ \ r>>R_V\nonumber\\
    &\alpha_{cl} >> 1, \ \ \ \ \ \ \alpha_q >>1 \ \ \ \ \ \ \ \ \text{when} \ \ \ \ \ r<< \dfrac{1}{E}\nonumber\\
    &\alpha_{cl}>>1, \ \ \ \ \ \ \alpha_q << 1 \ \ \ \ \ \ \ \ \text{when} \ \ \ \ \ \dfrac{1}{E} << r << R_V
\end{align}
We can see that there is a regime $\dfrac{1}{E} << r << R_V$ where although the classical non-linearities dominate over kinetic terms quantum corrections are still under control. On large scales both the parameters are negligible implying that classical non-linearities as well as quantum corrections loose dominance allowing us to bring about sought after modifications to gravity.

\section{Conclusion}\label{SecVI}
    Proca theories and its extension to curved spacetimes are the most general theories that attempt to modify gravity at large scales. These contain useful non-linear interactions in a way that keeps the theory classically sound by allowing only three degrees of freedom to propogate. It is expected that the theory is non-renormalisable, the situation being no better than that of GR. As a result it is only an effective field theory which makes it necessary to investigate its stability on quantum level. This becomes crucial since the theory relies on screening mechanism which comes into play at scales where quantum corrections can potentially dominate thereby destroying the ghost free classical structure.
    
    In this paper we investigated quantum stability of a restricted class of Proca theory in curved spacetime. One way to do this is to compute effective action which we did at one loop level up to second order in background vector field by employing the Vilkovisky-DeWitt prescription. Owing to the complexities with the extended Schwinger-DeWitt method we used a more approachable perturbative method which allows computation of effective action up to any desired order of background field. This gave us an off-shell effective action which is both gauge invariant as well gauge condition independent. As expected we found corrections involving higher order derivatives potentially leading to ghosts. In the limit of vanishing Proca mass we found that the most dangerous term came along with a coefficient $\kappa^2/m^2$. Demanding quantum corrections to remain suppressed below a UV scale $E$ where cubic interactions become important we found a range \hyperref[bbound]{(70)} within which the coupling constant should lie. The best choice among the values within this range is that which maximises the energy scale which we found to be $b,c_1 \sim \sqrt{m^3\kappa^3}$ giving the energy scale $E = \sqrt{mM_P}$. This in turn gave us the decoupling limit \hyperref[dl1]{(66)} in which the transverse and longitudinal modes of vector field and $h_{\mu\nu}$ completely decouple from each other.
    
    The quantum corrected action \hyperref[qcaction]{(71)} features a classical non-linear term and the quantum corrections supprressed by energy scale $E$ giving rise to the possibility that quantum corrections could dominate at scale where classical non-linear term becomes important compared to kinetic terms. A straightforward comparison of these terms with the kinetic terms however gave us distinct expansion parameters \hyperref[param]{(73)}: $\alpha_{cl}$ controlling the classical non-linearity and $\alpha_q$ controlling the quantum corrections thereby establishing an hierarchy among them. Furthermore owing to similar structure of action in decoupling limit as well the classical and quantum expansion parameters as compared to previous works we found that there exists a regime where classical non-linearty dominates while quantum corrections are still suppressed protecting EFT at scales where Vainshtein mechanism can play its role in screening the extra degree of freedom.
    
   VDEA furnishes quantum corrections which are gauge invariant and gauge condition independent which is necessary especially when we want them to be off-shell in gauge theories. It would certainly be interesting to go a step further by including the $\mathcal{L}_5$ and $\mathcal{L}_6$ Lagrangians. However as already shown we need to face lengthy expressions and extract the relevant loop integrals out of it if we use the perturbative technique (which is more approachable compared to generalised Schwinger DeWitt method) and these are immense in number especially when gravity is treated as a dynamical field. In light of this, we leave the computation of effective action up to quadratic order (or even beyond) in general background vector field using the Vilkovisky- DeWitt prescription and subsequent stability analysis including the $\mathcal{L}_5$ and $\mathcal{L}_6$ Lagrangians for future work.
     \section*{Acknowledgement}
	The authors of this paper would like to thank Prof. Lavinia Heisenberg for her comments on our previous paper which led us to this work. The calculations in this paper have been carried out in MATHEMATICA using the xAct packages xTensor \cite{xtensor} and xPert \cite{xpert}. This work is partially supported by DST (Govt. of India) Grant
    No. SERB/PHY/2017041.

\appendix
\section{Field space metric and connections}\label{AppA}
The inverse field space metric is given by
\begin{align}\label{invm}
&(G^{11})_{\mu\nu\alpha\beta}(x,x') = \delta(x,x')\Big\{\Big(1+\dfrac{c_1 A^2}{4}\Big)\mathcal{G}_{\mu\nu\alpha\beta} - c_1\Big(\dfrac{c_1}{C}-1\Big)A_{(\mu} g_{\nu)(\beta}A_{\alpha)} + \dfrac{c_1}{4}(A_\mu A_\nu g_{\alpha\beta} + A_\alpha A_\beta g_{\mu\nu})\Big\}\nonumber\\
&(G^{12})_{\alpha\beta\mu} = \delta(x,x')\dfrac{c_1 A_{(\alpha}g_{\beta)\mu}}{C}\nonumber\\
&(G^{22})_{\alpha\beta}(x,x') = \delta(x,x')\Big\{\dfrac{g_{\alpha\beta}}{C}\Big(1 + \dfrac{c_1^2}{C}A^2\Big) -\dfrac{c_1^2 A_\alpha A_\beta}{2C^2}\Big\}
\end{align}
Using these one can compute the Christoffel connections
\begin{align}\label{chrisconn}
 &(\Gamma^1_{11})^{\rho\sigma\mu\nu}_{\lambda\tau}(x,x',x'') = \tilde{\delta}(x'',x')\tilde{\delta}(x'',x)\Big\{-\delta^{(\mu}_{(\lambda}g^{\nu)(c_1}\delta^{\sigma)}_{\tau)} +\dfrac{1}{4}g^{\mu\nu}\delta^{\rho}_{(\lambda}\delta^{\sigma}_{\tau)}+\dfrac{1}{4}g^{\rho\sigma}\delta^{\mu}_{(\lambda}\delta^{\nu}_{\tau)} +\dfrac{1}{4}g_{\lambda\tau}g^{\mu(\rho}g^{\sigma)\nu} -\dfrac{1}{8}g_{\lambda\tau}g^{\mu\nu}g^{\rho\sigma}\nonumber\\
 &\hspace{30mm} +\dfrac{c_1}{2}A^{(\rho}\delta^{\sigma)}_{(\lambda}A^{(\mu}\delta^{\nu)}_{\tau)} + \dfrac{c_1}{4}A^{(\rho}g^{\sigma)(\mu}A^{\nu)}g_{\lambda\tau}-\dfrac{c_1}{16}A^2g_{\lambda\tau}g^{\rho(\mu}g^{\nu)\sigma} + \dfrac{c_1}{32}A^2g_{\lambda\tau}g^{\mu\nu}g^{\rho\sigma}\nonumber\\
 &\hspace{30mm} + \dfrac{c_1^2}{4C}(A_{(\lambda}A^{(\mu}\delta^{\nu)}_{\tau)}g^{\rho\sigma} + A_{(\lambda}A^{(\rho}\delta^{\sigma)}_{\tau)}g^{\mu\nu}) - \dfrac{c_1^2}{2C}(A_{(\lambda}\delta^{(\rho}_{\tau)}g^{\sigma)(\mu}A^{\nu)} + A_{(\lambda}\delta^{(\mu}_{\tau)}g^{\nu)(\rho}A^{\sigma)})\Big\}\nonumber\\
 &(\Gamma^1_{12})^{\rho\sigma\mu}_{\lambda\tau}(x,x',x'') = \dfrac{\tilde{\delta}(x'',x')\tilde{\delta}(x'',x)c_1}{4}\Big\{-A^\mu\delta^{(\rho}_{\lambda}\delta^{\sigma)}_{\tau} - A_{(\lambda}\delta^{\mu}_{\tau)}g^{\rho\sigma} + 2A_{(\lambda}\delta^{(\rho}_{\tau)}g^{\sigma)\mu} - 2A^{(\rho}g^{\sigma)\mu}g_{\lambda\tau} + A^\mu g_{\lambda\tau}g^{\rho\sigma}\Big\}\nonumber\\
 &(\Gamma^1_{22})^{\rho\sigma}_{\lambda\tau}(x,x',x'') = \dfrac{\tilde{\delta}(x'',x')\tilde{\delta}(x'',x)}{4}\Big\{\Big(\dfrac{C-2c_1}{2}\Big)\Big(1 + \dfrac{c_1 A^2}{4}\Big)\delta^{(\rho}_{\lambda}\delta^{\sigma)}_{\tau} + \dfrac{c_1}{2}\Big(3c_1 - C - \dfrac{2c_1^2}{C}\Big)A_{(\lambda}\delta^{(\rho}_{\tau)}A^{\sigma)}\nonumber\\
 &\hspace{30mm} -\dfrac{c_1(c_1 -C)}{8}A^\rho A^\sigma g_{\lambda\tau} + \dfrac{c_1}{8}\Big(C - \dfrac{4c_1(C-2c_1)}{C}\Big)A_\lambda A_\tau g^{\rho\sigma} + \dfrac{c_1(2c_1-C)}{16}A^2g_{\lambda\tau}g^{\rho\sigma}\Big\}\nonumber\\
 &(\Gamma^2_{11})^{\rho\sigma\mu\nu}_{\lambda}(x,x',x'')=\dfrac{\tilde{\delta}(x'',x')\tilde{\delta}(x'',x)c_1}{4C}\Big\{A^{(\rho}\delta^{\sigma)}_{\lambda}g^{\mu\nu} + A^{(\mu}\delta^{\nu)}_{\lambda}g^{\rho\sigma} - 2A^{(\rho}g^{\sigma)(\mu}\delta^{\nu)}_{\lambda} + 2A^{(\mu}g^{\nu)(\rho}\delta^{\sigma)}_{\lambda}\Big\}\nonumber\\
 &(\Gamma^2_{12})^{\rho\sigma\mu}_{\lambda}(x,x',x'') = \dfrac{\tilde{\delta}(x'',x')\tilde{\delta}(x'',x)}{4}\Big\{\Big(1-\dfrac{c_1^2A^2}{2C}\Big)(\delta^\mu_{\lambda} g^{\rho\sigma} - 2\delta^{(\rho}_{\lambda} g^{\sigma)\mu}) - \dfrac{c_1^2}{C}A^\mu A^{(\rho}\delta^{\sigma)}_{\lambda} +\dfrac{3c_1^2}{C}A_\lambda A^{(\rho}g^{\sigma)\mu} - \dfrac{c_1^2}{2C}A_\lambda A^\mu g^{\rho\sigma}\Big\}\nonumber\\
 &(\Gamma^2_{22})^{\rho\sigma}_{\lambda}(x,x',x'')=\dfrac{\tilde{\delta}(x'',x')\tilde{\delta}(x'',x)c_1}{4C}\{2(C-2c_1)A^{(\rho}\delta^{\sigma)}_{\lambda}-(C-4c_1)A_\lambda g^{\rho\sigma}\}
 \end{align}
 where $C = 1 - 2c_1c_2$
\section{Loop Integrals}\label{AppB}
    We encounter the following loop integrals while computing correlators whose divergent part is presented below,
    \begin{align}
        &\int d^4k\dfrac{1}{m^2 + Ck^2} = \dfrac{-m^2}{8\pi^2C^2\epsilon}\nonumber\\
        &\int d^4k\dfrac{k_\mu k_\nu}{m^2 + Ck^2} = \dfrac{\eta_{\mu\nu}m^4}{32\pi^2C^3\epsilon}\nonumber\\
        &\int d^4k\dfrac{k_\mu k_\nu k_\alpha k_\beta}{m^2 + Ck^2} = -\dfrac{m^6\text{Perm}[\eta_{\mu\nu}\eta_{\alpha\beta}]}{192\pi^2C^4\epsilon}\\\nonumber\\
         &\int d^4p \dfrac{1}{m^2 + C(k-p)^2} = -\dfrac{m^2}{8\pi^2C^2\epsilon}\nonumber\\
         &\int d^4p \dfrac{p_\mu}{m^2 + C(k-p)^2} = \dfrac{-k_\mu m^2}{8\pi^2C^2\epsilon}\nonumber\\
         &\int d^4p \dfrac{p_\mu p_\nu}{m^2 + C(k-p)^2} = \dfrac{-k_\mu k_\nu m^2}{8\pi^2 c^2\epsilon} + \dfrac{\eta_{\mu\nu}m^4}{32\pi^2C^3\epsilon}\nonumber\\
         &\int d^4p \dfrac{p_\mu p_\nu p_\alpha}{m^2 + C(k-p)^2} = -\dfrac{k_\mu k_\nu k_\alpha m^2}{8\pi^2C^2\epsilon} + \dfrac{m^4\text{Perm}[\eta_{\mu\nu}k_\alpha ]}{32\pi^2C^3\epsilon}\nonumber\\
         &\int d^4p \dfrac{p_\mu p_\nu p_\alpha p_\beta}{m^2 + C(k-p)^2} = -\dfrac{k_\mu k_\nu k_\alpha k_\beta m^2}{8\pi^2 C^2\epsilon} + \dfrac{m^4\text{Perm}[\eta_{\mu\nu}k_\alpha k_\beta]}{32\pi^2C^3\epsilon} \nonumber - \dfrac{m^6\text{Perm}[\eta_{\mu\nu}\eta_{\alpha\beta} ]}{192\pi^2C^4\epsilon}\nonumber\\
         &\int d^4p \dfrac{1}{p^2(m^2 + C(k-p)^2)} = \dfrac{1}{8\pi^2C\epsilon}\nonumber\\
         &\int d^4p \dfrac{p_\mu}{p^2(m^2 + C(k-p)^2)} = \dfrac{k_\mu m^2}{16\pi^2C\epsilon}\nonumber\\
         &\int d^4p \dfrac{p_\mu p_\nu}{p^2(m^2 + C(k-p)^2)} = \dfrac{k_\mu k_\nu m^2}{24\pi^2 C\epsilon} - \dfrac{\eta_{\mu\nu}k^2}{96\pi^2C\epsilon}-\dfrac{\eta_{\nu\nu}m^2}{32\pi^2C^2\epsilon}\nonumber\\
         &\int d^4p \dfrac{p_\mu p_\nu p_\alpha}{p^2(m^2 + C(k-p)^2)} = \dfrac{k_\mu k_\nu k_\alpha m^2}{32\pi^2C\epsilon} - \Big(k^2 + \dfrac{4m^2}{C}\Big)\dfrac{\text{Perm}[\eta_{\mu\nu}k_\alpha ]}{192\pi^2C\epsilon}\nonumber\\
        &\int d^4p \dfrac{p_\mu p_\nu}{p^4(m^2 + C(k-p)^2)} =  \dfrac{\eta_{\mu\nu}}{32\pi^2C\epsilon} \nonumber\\
         &\int d^4p \dfrac{p_\mu p_\nu p_\alpha}{p^4(m^2 + C(k-p)^2)} = \dfrac{\text{Perm}[\eta_{\mu\nu}k_\alpha ]}{96\pi^2C\epsilon}\nonumber\\
         &\int d^4p \dfrac{p_\mu p_\nu p_\alpha p_\beta}{p^4(m^2 + C(k-p)^2)} =  \dfrac{\text{Perm}[\eta_{\mu\nu}k_\alpha k_\beta]}{192\pi^2C\epsilon}- \Big(k^2 + \dfrac{2m^2}{C}\Big)\dfrac{\text{Perm}[\eta_{\mu\nu}\eta_{\alpha\beta} ]}{384\pi^2C\epsilon}
    \end{align}
    \begin{align}
         &\int d^4p \dfrac{1}{(m^2 + Cp^2)(m^2 + C(k-p)^2)} = \dfrac{1}{8\pi^2C^2\epsilon}\nonumber\\
         &\int d^4p \dfrac{p_\mu}{(m^2 + Cp^2)(m^2 + C(k-p)^2)} = \dfrac{k_\mu}{16\pi^2C^2\epsilon}\nonumber\\
         &\int d^4p \dfrac{p_\mu p_\nu}{(m^2 + Cp^2)(m^2 + C(k-p)^2)} = \dfrac{k_\mu k_\nu}{24\pi^2C^2\epsilon} -\Big(k^2 + \dfrac{6m^2}{C}\Big)\dfrac{\eta_{\mu\nu}}{96\pi^2C^2\epsilon}\nonumber\\
         &\int d^4p \dfrac{p_\mu p_\nu p_\alpha}{(m^2 + Cp^2)(m^2 + C(k-p)^2)} = \dfrac{k_\mu k_\nu k_\alpha}{32\pi^2C^2\epsilon} - \Big(k^2 + \dfrac{6m^2}{C}\Big)\dfrac{\text{Perm}[k_\mu \eta_{\nu\alpha}]}{192\pi^2C^2\epsilon}\nonumber\\
         &\int d^4p \dfrac{p_\mu p_\nu p_\alpha p_\beta}{(m^2 + Cp^2)(m^2 + C(k-p)^2)} =\dfrac{k_\mu k_\nu k_\alpha k_\beta}{40\pi^2C^2\epsilon} - \Big(\dfrac{k^2}{20} + \dfrac{m^2}{3C}\Big)\dfrac{\text{Perm}[\eta_{\mu\nu}k_\alpha k_\beta]}{16\pi^2C^2\epsilon} + \Big(\dfrac{m^4}{C^2} + \dfrac{k^2m^2}{3C} + \dfrac{k^4}{30}\Big)\dfrac{\text{Perm}[\eta_{\mu\nu}\eta_{\alpha\beta}]}{64\pi^2C^2\epsilon}\nonumber\\
         &\int d^4p \dfrac{p_\mu p_\nu p_\alpha p_\beta p_\lambda}{(m^2 + Cp^2)(m^2 + C(k-p)^2)} =\dfrac{k_\mu k_\nu k_\alpha k_\beta k_\lambda}{48\pi^2C^2\epsilon} - \Big(\dfrac{m^2}{4C} + \dfrac{k^2}{30}\Big)\dfrac{\text{Perm}[\eta_{\mu\nu}k_\alpha k_\beta k_\lambda]}{16\pi^2C^2\epsilon} \nonumber\\&\hspace{60mm}+ \Big(\dfrac{m^4}{2C^2} + \dfrac{m^2k^2}{6C} + \dfrac{k^4}{60}\Big)\dfrac{\text{Perm}[k_\lambda\eta_{\mu\nu}\eta_{\alpha\beta}]}{64\pi^2C^2\epsilon}
     \end{align}
     where $C = 1-2c_1c_2$ and $\epsilon = 4 - n$, \ $n$ being spacetime dimension. The function "Perm" instructs to take all possible distinct permutations of indices. For ex:
     \begin{equation}\nonumber
        \text{Perm}[\eta_{\mu\nu}k_\alpha] = \eta_{\mu\nu}k_\alpha + \eta_{\mu\alpha}k_ \nu + \eta_{\nu\alpha}k_\mu
    \end{equation}
\section{Divergences in one loop effective action}\label{AppC}
The functions appearing in the divergent part of one loop effective action \hyperref[diveffac1]{(48)} are as follows
 \begin{align}
     &f_1 = \dfrac{1}{C^3L}\Big\{3m^2b^2 + \kappa^2\Big(\dfrac{3m^4c_1C}{4} - \dfrac{3m^4c_1^2}{2} + \dfrac{5m^4C^2}{32}\Big)\Big\}\nonumber\\
     &f_2 = -\dfrac{3m^2b}{C^2L}\nonumber\\
     &f_3 = \dfrac{1}{C^2L}\Big\{11b^2 - \kappa^2\Big(\dfrac{35m^2C^2}{96} + \dfrac{15m^2c_1C}{8} + 3m^2c_1^2\Big)\Big\}\nonumber\\
     &f_4 = \dfrac{1}{C^2L}\Big\{8b^2 + \kappa^2\Big(\dfrac{m^2C^2}{3} + \dfrac{15m^2c_1C}{8} + 3m^2c_1^2\Big)\Big\}\nonumber\\
     &f_5 = -\dfrac{1}{L}\Big\{\dfrac{9b^2}{5m^2C} + \dfrac{49\kappa^2C}{24}\Big\}\nonumber\\
     &f_6 = -\dfrac{1}{L}\Big\{\dfrac{53b^2}{15m^2C} - \dfrac{49\kappa^2C}{24}\Big\}\nonumber\\
     &f_7 = \dfrac{7\kappa^2C^2}{48m^2L}\nonumber\\
     &f_8 = \dfrac{1}{L}\Big\{\dfrac{4b^2}{15m^4} -  \dfrac{7\kappa^2C^2}{48m^2}\Big\}
\end{align}
 \hspace{25mm} where \ \ $L = \dfrac{1}{16\pi^2\epsilon} \ \ \ \ \text{and} \ \ \ \ \ C = 1 - 2c_1c_2$.
 
 The coefficients of the non-gauge invariant and the gauge invariant structures appearing in \hyperref[diveffac2]{(49)} are as follows
 \begin{align}
     &\tilde g_1 = \dfrac{1}{C^3L}\Big\{3m^2b^2 + \kappa^2\Big(\dfrac{3m^4c_1C}{4} - \dfrac{3m^4c_1^2}{2} + \dfrac{5m^4C^2}{32}\Big)\Big\}\nonumber\\
     &\tilde g_2 = -\dfrac{3m^2b}{C^2L}\nonumber\\
     &\tilde g_3 = \dfrac{1}{C^2L}\Big\{11b^2 - \dfrac{35\kappa^2m^2C^2}{96}\Big\}\nonumber\\
     &\tilde g_4 = \dfrac{1}{C^2L}\Big\{8b^2 + \dfrac{\kappa^2m^2C^2}{3}\Big\}\nonumber\\
     &\tilde g_5 = -\dfrac{9b^2}{5m^2CL}\nonumber\\
     &\tilde g_6 = -\dfrac{53b^2}{15m^2CL}\nonumber\\
     &\tilde g_8 = \dfrac{4b^2}{15m^4L}\nonumber\\
     &g_1 = \dfrac{15\kappa^2m^2c_1}{16CL} + \dfrac{3\kappa^2m^2c_1^2}{2C^2L}\nonumber\\
     &g_2 = \dfrac{49\kappa^2C}{48L}\nonumber\\
     &g_3 = -\dfrac{7\kappa^2C^2}{96m^2L}
 \end{align}
 
\end{document}